\documentstyle[12pt]{article}
\newif\ifall\alltrue
\def\private#1{}			
\def\eps{\varepsilon}

\def\ff{\Longrightarrow}

\def\nq{\hspace{-1em}}

\def\ignore#1{}
\def\deltabar{\delta\!\!\!^{-}\,}
\def\hbar{h\!\!\!\!^{-}\,}
\def\dbar{d\!\!^{-}\!}
\def\beq{\begin{equation}}
\def\eeq{\end{equation}}
\def\bqa{\begin{eqnarray}}
\def\eqa{\end{eqnarray}}

\begin{document}
\ifall

\begin{titlepage}
\hspace*{13cm}LMU 95-..

\hspace*{13cm}May 1995

\begin{center}					  \vspace*{3cm}
{\LARGE Proton Spin in the                       }\\[0.5cm]
{\LARGE  Instanton Background                    }\\[3cm]
  {\bf Marcus Hutter\footnotemark}                \\[2cm]
  {\it Sektion Physik der Universit\"at M\"unchen}\\
  {\it Theoretische Physik}                       \\
  {\it Theresienstr. 37 $\quad$ 80333 M\"unchen} \\[2cm]
\end{center}
\footnotetext{E--Mail:hutter@hep.physik.uni--muenchen.de}

\begin{abstract}
The proton form factors are reduced to vacuum correlators of 4 quark
fields by assuming independent constituent quarks.
The axial singlet quark and gluonic form factors are calculated
in the instanton liquid model.
A discussion of gauge(in)dependence is given.
\end{abstract}
\end{titlepage}

{\parskip=0.2ex\tableofcontents}

\fi
\unitlength=1.00mm

\section{Introduction}\label{sec61}
A variety of predictions concerning chiral symmetry breaking
can be made within the instanton liquid model. Although the
't Hooft interaction \cite{tHo} explicitly breaks the $U(1)$
axial symmetry, instanton models are up to now not too successful
in describing quantitatively the axial singlet channel.
The most interesting quantities are the $\eta'$ mass
and the spin of the proton.

Sections \ref{sec62}, \ref{sec63} and \ref{sec64} are an introduction
to the proton spin problem. In section \ref{sec65} the
proton form factors are reduced to vacuum correlators of 4 quark
fields by assuming independent constituent quarks.
The axial singlet quark and gluonic form factors are calculated
in section \ref{sec66}, \ref{sec67} and \ref{sec68} by using
the propagator and 4 point functions of the instanton liquid model.
Gauge(in)dependence is examined. A discussion
of the results and a comparison with \cite{For} is given in
section \ref{sec69}.

\section{Measurement of the Axial Form Factors}\label{sec62}

The forward matrix elements of the axial currents
$$
  s_\mu\Delta\psi = \langle ps|\bar\psi\gamma_\mu\gamma_5\psi
  |ps\rangle \quad,\quad \psi = u,d,s
$$
can be interpreted as the quark spin content of the proton,
in a sense defined more accurately in the following sections.
Three independent linear combinations of
$\Delta\psi$ have been measured, thus allowing to extract their
individual values.

{}From the neutron $\beta$-decay, using isospin invariance,
one gets \cite{Sehgal,Jaf}
$$
  a_3 = g_A = \Delta u-\Delta d = F+D = 1.254\pm 0.06 \quad.
$$
{}From the octet hyperon $\beta$-decay, using $SU(3)_F$ symmetry,
one gets \cite{Bourquin,Jaf}
$$
  \sqrt{3}a_8 = \Delta u+\Delta d -2\Delta s = 3F-D = 0.688\pm 0.0035
  \quad.
$$
{}From the spin dependent structure function $g_1^p$ of the proton,
which has been measured by EMC \cite{EMC} and SMC \cite{SMC}, one
can extract
\beq
  \Gamma_p=\int_0^1\!g_1^p(x)dx =
  {4\over 9}\Delta u+{1\over 9}\Delta d+{1\over 9}\Delta s+O(\alpha_s)
  = 0.142\pm0.014
\eeq
where we have given the world average value.

Of special interest is the quarkspin sum, which can be extracted
from the values given above,
\beq
  \Delta\Sigma^{GI}=\sqrt{3\over 2}a_0 = \Delta u+\Delta d+\Delta s =
  0.27\pm 0.13
\eeq
where the $O(\alpha_s)$ corrections have been included. It deviates
significantly from the naive quark model value $\Delta\Sigma_{qm}=1$.
This deviation is the origin of the so called spin problem.
Further the large polarization of strange quarks in the proton
$$
  \Delta s = -0.1\pm 0.05
$$
is counter intuitive because this indicates a large strange quark
content of the proton.

Much more could be said about proton spin phenomenology and
the experiments. For an introduction and further references see
\cite{Baas,Reya,Roberts,EMC}.
We will now give a more thorough definition and
interpretation of $\Delta\Sigma^{GI}$ and other
quantities, which we want to calculate within the instanton model.

\section{Axial Singlet Currents \& Anomaly}\label{sec63}

It is well known that products of operators at the same spacetime
point are very singular objects. In order to make the expressions
well defined one has to regularize and renormalize the operator
products. An anomaly appears, if this procedure breaks a symmetry
of the theory. The most important ones are the breakdown of
the scale invariance and the breakdown of the axial symmetry \cite{Shi}.
In the following we are interested in the axial anomaly \cite{ABJ}.
The operator product which has to be regularized is the axial
singlet current,
\beq
  J_{\mu 5}(x) = \sum_{q\in\{u,d,s,\ldots\}}
                 \bar{q}(x)\gamma_\mu\gamma_5 q(x)
\eeq
which seems to be local, gauge invariant and conserved\footnote{
We will use the term 'conserved' even for $m_q\neq 0$. Sometimes
this current is called the symmetric current in the literature.}.
Unfortunately after regularization one of the three properties
is unavoidably lost. Therefore we can define two different local
currents, a conserved (c) one and a gauge invariant (GI) one.
The third GI, conserved and
non-local current is discussed in \cite{Kogut} in connection with
the $U(1)$ problem. We will suppress the summation over quark flavors
and write $\psi$ for the quark field operator:
\bqa
  J_{\mu 5}^{GI}(x) &=& \lim_{\eps\to 0}
    \bar\psi(x+\eps)\gamma_\mu\gamma_5
    P\exp\left(i\int_x^{x+\eps}\!\! dz\!\cdot\!A(z)\right)
    \psi(x) \nonumber\\
  J_{\mu 5}^c(x) &=& \lim_{\eps\to 0}
    \bar\psi(x+\eps)\gamma_\mu\gamma_5\psi(x)
\eqa
The difference between the two currents is described by the anomaly
current $K_\mu$:
\bqa\label{pro3}
  K_\mu(x) &=& {N_f\alpha_s\over 2\pi}\eps_{\mu\nu\rho\sigma}
    \mbox{tr}_c A^\nu(G^{\rho\sigma}-{2\over 3}A^\rho A^\sigma)
    \quad,\quad
    J_{\mu 5}^{GI} = J_{\mu 5}^c + K_\mu \nonumber\\
  \partial^\mu K_\mu(x) &=& {N_f\alpha_s\over 2\pi}\mbox{tr}_c G\tilde G(x)
    = a(x) \\
  \partial^\mu J_{\mu 5}^c(x) &=& 2mJ_5(x) \quad\quad\quad,\quad
    J_5 = i\bar\psi\gamma_5\psi \quad. \nonumber
\eqa
$m$ is the current quark mass and $N_f$ is the number of quark flavors.
Note, that the splitting of
$J_{\mu 5}$ in a conserved and an anomaly part is gauge dependent.
There are attempts to define both uniquely on physical
grounds \cite{Reya}. The intention is to define $J_{\mu 5}^c$ as the
naive parton model spin and $K_\mu$ as some gluonic contribution.
The proton matrix elements of the various currents can be expressed
in terms of real form factors $G_i, K_i, A and J$:
\bqa\label{pro4}
  \langle p' s'|J_{\mu 5}^{GI}(0)|ps\rangle &=&
    \bar u_{s'}(p')\left[\gamma_\mu\gamma_5
    G_1^{GI}(q^2)-q_\mu\gamma_5 G_2^{GI}(q^2)\right] u_s(p) \nonumber\\
  \langle p' s'|J_{\mu 5}^c(0)|ps\rangle &=&
    \bar u_{s'}(p')\left[\gamma_\mu\gamma_5
    G_1^c(q^2)-q_\mu\gamma_5 G_2^c(q^2)\right] u_s(p) \nonumber\\
  \langle p' s'|K_\mu(0)|ps\rangle &=&
    \bar u_{s'}(p')\left[\gamma_\mu\gamma_5
    K_1(q^2)-q_\mu\gamma_5 K_2(q^2)\right] u_s(p)              \\
  \langle p' s'|a(0)|ps\rangle &=&
    2MiA(q^2)\bar u_{s'}(p')\gamma_5 u_s(p) \nonumber\\
  \langle p' s'|J_5(0)|ps\rangle &=&
    iJ(q^2)\bar u_{s'}(p')\gamma_5 u_s(p) \nonumber
\eqa
$M$ is the proton mass and $q=p'-p$.
{}From (\ref{pro3}) one can derive the following
relations between the form factors:
$$
  G_1^{GI}=G_1^c+K_1 \quad,\quad G_2^{GI}=G_2^c+K_2
$$
\beq\label{pro5}
  G_1^c-{q^2\over 2M}G_2^c={m\over M}J \quad,\quad
  K_1-{q^2\over 2M}K_2 = A
\eeq
$$
  G_1^{GI}-{q^2\over 2M}G_2^{GI} = {m\over M}J+A \quad,
$$
where all form factors are evaluated at $q$.
The last equation relates only GI quantities.
In the next section we show that the form factor $G_1^{GI}$
at zero momentum transfer can
be connected with the proton spin.

\section{The Proton Spin and its Interpretation}\label{sec64}

The stress tensor $T_{\mu\nu}$ is conserved
$(\partial_\mu T^{\mu\nu}=0)$ symmetric and GI and can be
constructed from the Noether theorem. The angular momentum density
tensor $M^{\mu\nu\rho}$
associated with Lorentz transformations can be expressed
in terms of $T_{\mu\nu}$:
\beq
  M^{\mu\nu\rho} = x^\nu T^{\mu\rho} - x^\rho T^{\mu\nu}
\eeq
$M$ can be decomposed in spin and orbital contribution of quarks
and gluons \cite{Jaf}:
\bqa
  M^{\mu\nu\rho} &=&
    M^{\mu\nu\rho\nq\nq}_{q,orb} +
    M^{\mu\nu\rho\nq\nq}_{q,spin} +
    M^{\mu\nu\rho\nq\nq}_{g,orb} +
    M^{\mu\nu\rho\nq\nq}_{g,spin}
  - {1\over 4}G^2(x^\nu g^{\mu\rho}-x^\rho g^{\mu\nu})
  + \partial(\cdots)
  \nonumber\\
  M^{\mu\nu\rho\nq\nq}_{q,orb}  &=& {1\over 2}i\bar\psi
    \gamma^\mu(x^\nu\partial^\rho-x^\rho\partial^\nu)\psi
  \quad,\quad
  M^{\mu\nu\rho\nq\nq}_{q,spin} = {1\over 2}\eps^{\mu\nu\rho\sigma}
    \bar\psi\gamma_\sigma\gamma_5\psi = {1\over 2}J^{GI}\nq_{\sigma 5}
  \\
  M^{\mu\nu\rho\nq\nq}_{g,orb}  &=& -G^{\mu\sigma}
    (x^\nu\partial^\rho-x^\rho\partial^\nu)A_\sigma
  \quad,\quad
  M^{\mu\nu\rho\nq\nq}_{g,spin} =
    G^{\mu\rho}A^\nu-G^{\mu\nu}A^\rho
  \nonumber
\eqa
The last two terms in $M^{\mu\nu\rho}$ do not contribute to the
angular momentum operator
\beq\label{pro9}
  J^i = {1\over 2}\eps^{ijk}\int\!d^3\!x\; M^{0jk}(x) \quad.
\eeq
Taking the matrix element of $J_z$ in a proton state,
where the proton is
aligned in $z$-direction and at rest we
get the spin of the proton
\beq
  \Delta J = {1\over{\cal N}}\langle ps|J_z|ps\rangle
    = {1\over 2}\eps^{3jk}\langle ps|M^{0jk}(0)|ps\rangle
  \quad,\quad
    {\cal N} = \langle p,s|p,s\rangle = \deltabar^3(0)
\eeq
The total spin of the proton is with no doubt $1/2$ and we get
the sum rule
\beq
  \Delta J = \Delta L_q + {1\over 2}\Delta\Sigma^{GI} +
             \Delta L_g + \Delta g = {1\over 2}
\eeq
where $(\Delta L_q,{1\over 2}\Delta\Sigma^{GI},\Delta L_g,\Delta g)$
are the (quark-orbital, quark-spin, gluon-orbital, gluon-spin)
contribution to the proton spin, defined as matrix elements
of the various parts of $M$ given above.
Therefore The GI axial current measures the quark spin
contribution to the proton spin. The space integral in \ref{pro9} can be
cancelt with the state normalization and we get in covariant notation:
\beq\label{pro7}
  s_\mu\Delta\Sigma^{GI} = \langle ps|J_{\mu 5}^{GI}(0)|ps\rangle
  \quad\ff\quad \Delta\Sigma^{GI} = G_1^{GI}(0)
\eeq
In the naive quark model the proton consists of three quarks
at rest. There is no orbital and no gluonic contribution to
the proton spin. This leads to the Ellis-Jaffe sum rule
$\Delta J$ = ${1\over 2}\Delta\Sigma^{GI}$ = $1/2$.
In the real world the identification of $\Delta\Sigma^{GI}$ with
the proton spin is not correct, because $J_{\mu 5}^{GI}$
measures the spin of the (nearly massless) current quarks
whereas the proton consists of three massive ($\approx 300\;$MeV)
constituent quarks. Further in a model of non-interacting
constituent quarks the axial current which measures the
constituent quark spin should be anomaly free because
the anomaly is due to the interaction with gluons.
Therefore the conserved current $J_{\mu 5}^c$ might be identified
with the constituent quark spin operator.
\beq\label{pro11}
  s_\mu\Delta\Sigma^c = \langle ps|J_{\mu 5}^c(0)|ps\rangle
  \quad\ff\quad \Delta\Sigma^c = G_1^c(0) \stackrel{?}{=}1 \quad.
\eeq
{}From (\ref{pro5}) we get
\beq\label{pro12}
  \Delta\Sigma^{GI} = \Delta\Sigma^c + K_1(0)
\eeq
which can now be interpreted in the following way: The spin
of the constituent quarks $\Delta\Sigma^c$ are formed by the
spin of the current quarks $\Delta\Sigma^{GI}$ and a rest
$-K_1(0)$, which contains orbital and gluonic contributions.
The origin of these contributions is {\it not} the motion
and interaction of the constituent quarks inside the proton,
because the constituent quarks are noninteracting and at rest
in the naive quark model, but
due to the formation of massive quarks from massless quarks.
Therefore (\ref{pro12}) may be discussed for an individual
''constituent'' quark.
Further the gluonic configurations which are responsible
for the generation  of the quarkmass also determine the value of
$K_1(0)$.

E.g.\ in a BAG model a massive quark is formed by confining a
massless quark to a sphere. The spin of the massive constituent quark
is the sum of the spin $({1\over 2}\Delta\Sigma^{GI})$
and the oribtal ${1\over 2}(1-\Delta\Sigma^{GI})$ contribution of
the current quark. The BAG, which might be formed by nonperturbative
gluonic configurations, is responsible for the mass generation
and indirectly for the orbital contribution. From analytical and
numerical calculations we know, that in the BAG model
the constituent spin
is splitted into $70\%$ spin and $30\%$ orbital contribution when
starting with massless quarks. 

Whereas (\ref{pro12}) is rigorously true, the interpretation of ${1\over
2}\Delta\Sigma^c$
as the spin of a constituent quark and its value ${1\over 2}$
is questionable. One reason is, that
an axial current which describes massive constituent quarks
is by no means conserved in contradiction to $J_{\mu 5}^c$.

There exists another relation between $\Delta\Sigma^{GI}$ and
the form factor $A$ at zero momentum transfer. Before deriving
this relation we have to give a short discussion about
the order of limits and massless poles.
The following limits are taken: The spacetime volume goes to
infinity $(V_4\to\infty)$, because the universe is actually very large,
the current quark masses go to zero ($m\to 0)$, because the
up and down masses are very small and $q\to 0$, because we
are interested in the forward matrix elements. In principle the
results can depend on the order of the limits and therefore they
have to be choosen consistent with the physical situation. This
means, that if in the real world e.g. $q\ll m$ we first have to
take $q\to 0$ and then $m\to 0$. Actually we are interested in
the forward matrix element ($q\equiv 0$) and $m\neq 0$ in the real
world and the order of limits just stated applies. Through the
cluster theorem connected correlators in coordinate space
have to decay  to zero
when the separation of two arguments tends to infinity. Therefore
there are no $\delta(q)$-peaks in momentum space and the order of limits
$q\to 0$ and $V_4\to\infty$ can be taken at will. Because $m^4V_4\gg 1$
we have to take first $V_4\to\infty$ and then $m\to 0$.
In statistical physics this is a well known fact, that
a spontaneous breakdown of a symmetry only occurs, when there is
a small explicit symmetry breaking term and the system volume tends to
infinity. In the final end one may remove the symmetry breaking term.
In QCD chiral symmetry is spontaneously broken (SBCS)
and the small current quark
mass is the explicit breaking of the chiral symmetry. Therefore
it is mandatory first to take $V_4\to\infty$ and then $m\to 0$
\cite{Smilga}. Therefore we can use the following order of limits
\beq
  \lim_{m\to 0}\{\lim_{q\to 0}[\lim_{V_4\to\infty}(\ldots)]\}.
\eeq
This justifies the usage of the infinite volume formulation from
the very beginning.

In real QCD there are no massless particles
$(m_\pi\neq 0)$. Therefore GI form factors have no massless poles
especially
\beq\label{pro13}
   q^2G_2^{GI}(q^2)\stackrel{q\to\ 0}{\longrightarrow}0
\eeq
{}From (\ref{pro5}), (\ref{pro7}) and (\ref{pro13}) we get
\beq\label{pro14}
  \Delta\Sigma^{GI} = {m\over M}J(0)+A(0)
\eeq
This relation is true wether there are Goldstone bosons in the
axial singlet channel or not. Experimentally we know that the lightest
particle in this channel is the $\eta'$ with a  mass
of 958 MeV much too large to be a Goldstone boson. Therefore $J(0)$
remains finite in the chiral limit and we obtain
\beq\label{pro20}
  \Delta\Sigma^{GI} = A(0) \quad\mbox{for}\quad m\to 0
\eeq
Assuming the non-existence of the axial singlet Goldstone boson
from the very beginning the order of limits is of no importance
in deriving (\ref{pro20}).
Note, that (\ref{pro20}) is only true,
if we take $m=m_u=m_d=m_s$, although all three masses tend to zero.
Otherwise additional nonsinglet currents on the r.h.s. of (\ref{pro14})
would survive the chiral limit \cite{Fri}.

Combining (\ref{pro5}), (\ref{pro13}) and (\ref{pro12})
we can conclude that
$$
  2M\Delta\Sigma^c = q^2G_2^c(q^2)_{|q^2=0} =
    -q^2K_2(q^2)_{|q^2=0}
$$
$\Delta\Sigma^c$ is given by the pole residuum
of $G_2^c$. Because $G_2^{GI}$ has no massless pole $\Delta\Sigma^c$ is
also given by the pole of $-K_2$. These massless poles are called
ghost poles and they may truly appear, even if there are no physical
massless particles, because $G_2^c$ and $K_2$ are gauge dependent
objects. Note that all other form factors defined in (\ref{pro5}) are GI
and therefore free of massless poles.

Table \ref{tabpro3} summarizes the values for the form factors
at zero momentum transfer for the following three cases:
\begin{itemize}\parskip=0ex\parsep=0ex\itemsep=0ex
\item the naive quark model of non-interacting constituent
      quarks of mass $m=M/N_f$,
\item chiral QCD and the identification of $\Delta\Sigma^c$ with
      the naive spin value 1,
\item the instanton liquid model.
\end{itemize}
In the following sections we will calculate some of the form factors
for a single constituent quark in the instanton liquid model.

\begin{table}[tbb]\label{tabpro3}
  \begin{center}\begin{tabular}{
         |c|c@{=}c@{+}c|c@{=}c@{+}c|c@{-}c@{=}c@{+}c@{=}c@{+}c|} \hline
    & $\Delta\Sigma^c$ & ${q^2\over 2M}G_2^c$ & ${m\over M}J$ &
      $K_1$ & ${q^2\over 2M}K_2$ & $A$ &
        $\Delta\Sigma^{GI}$ & ${q^2\over 2M}G_2^{GI}$ &
          $\Delta\Sigma^c$ & $K_1$ & ${m\over M}J$ & $A$     \\\hline
    $N_fm=M$ & 1 & 0 & 1 & 0 & 0 & 0 &
               1 & 0 & 1 & 0 & 1 & 0                         \\\hline
    $m=0$    & 1 & 1 & 0 & A-1 & -1 & A &
               A & 0 & 1 & A-1 &  0 & A                      \\\hline
    $Instanton$ & ? & ? & 0 & 0 & 1 & (-1) &
                  1 & 0 & ? & 0 & 0 & (-1)                   \\\hline
\end{tabular}\end{center}
\vspace{-3ex}
\caption[Proton form factors at zero momentum transfer]{
  \it The proton form factors at zero momentum transfer $q^2=0$
  in the naive constituent quark model ($N_fm=M$),
  in chiral QCD (m=0) and
  in the instanton-liquid model (Instanton).
  Experimentally $A$ is 0.27.
  }
\end{table}

\section{Reduction of the Proton Form Factors to Vacuum
Correlators}\label{sec65}

In this section we will calculate some of the form factors defined
above in the instanton liquid model. To apply the methods developed
in \cite{Hut2} we relate the form factors to vacuum correlation functions
\beq\label{pro29}
  \langle p's'|B(0)|ps\rangle =
\eeq
$$
   = -{1\over Z_\eta}\bar u_{s'}(p')\left[\int\!d^4\!x\,d^4\!z\;
    e^{ip'x-ipz}(i\partial\!\!\!/_x-M)(-i\partial\!\!\!/_z-M)
    \langle 0|{\cal T}\eta(x)B(0)\bar\eta(z)|0\rangle\right] u_s(p)
    \quad.
$$
$M$ is the proton mass and $B(0)$ is an arbitrary local operator.
$\eta(x)$ is a local operator with the
quantum numbers of a proton e.g. a product of three quark fields
in an appropriate spin and flavor combination \cite{Ioffe}.
Assuming that $\eta(x)$ tends to a free proton field operator for
infinit times the proton states can be reduced and (\ref{pro29})
is just an LSZ reduction formula for composite fields.
For our purpose the following form is more suitable
\bqa\label{pro30}
  \langle p's'|B(0)|ps\rangle &=&
    Z_\eta\bar u_{s'}(p')[\lim_{p^2,p'^2\to M^2}
    S^{-1}(p')T_B(p',p)S^{-1}(p)]u_s(p)             \nonumber\\
  T_B(p',p) &=& \int\!d^4\!x\,d^4\!z\;
    e^{ip'x-ipz}\langle 0|{\cal T}\eta(x)B(0)\bar\eta(z)|0\rangle    \\
  S(p) &=& \int\!d^4\!x\;
    e^{ipx}\langle 0|{\cal T}\eta(x)\bar\eta(0)|0\rangle =
    {iZ_\eta\over p\!\!/-M} + continuum       \nonumber\\
  Z_\eta^{1/2}u_s(p) &=& \langle 0|\eta(0)|ps\rangle    \nonumber
\eqa
The advantage of this form is, that the explicit knowledge of the
mass $M$ is not needed. In Euclidian calculations like lattice-,
instanton- and OPE-calculations it is always difficult to
extract pole masses.

This form can also be interpreted as a spectral
representation of the 3 point function. Inserting two complete
sets of states into the 3 point function
and taking the limit $p^2=p'^2\to M^2$ to select the proton state
one can directly attain (\ref{pro30}).

If e.g. $B(0)$ is a quark current, the 3 point function is a product
of 8 quark fields, which is too complicated to be evaluated in
a multi-instanton background. Let us assume that the proton consists
of three nearly independent quarks. Then the main nonperturbative
properties of the proton come from the formation of constituent
quarks out of current quarks. The forces which confine the constituent
quarks in the proton are assumed to modify the properties
of the proton only in a minor way, except that the proton is then stable.
This assumption is justified by the success of the constituent quark
model. The form factors of the proton are therefore the sum of the
form factors of the constituent quarks. $\eta$ has to be replaced
by a single quark field $\psi$ of flavor {\it up} or {\it down}
and $M$ must
be replaced by the constituent quark mass. In this case it is even more
important to use (\ref{pro30})
because one does not expect a definit pole mass
for the quark propagator. Looking at the quark propagator in
the instanton liquid model we see, that the $p\!\!/$ term remains
unrenormalized and therefore $Z_\psi=1$. For a constant constituent
mass this argument would be rigorously true. For a running mass it
is plausible that $Z_\psi$ is still approximately one. This
fact is true in all models of chiral symmetry breaking I know.
A conservative estimate is
\beq
  0.7\le Z_\psi\le 1
\eeq
In the following we will set $Z_\psi=1$ remembering that this not
an exact statement. The results for all form factors have to
be multiplied with $Z_\psi$.

\section{The Axial Form Factors $G_{1/2}^{GI}(q)$}\label{sec66}

The form factor of the current $j_\Gamma=\bar\psi\Gamma\psi$ of
a constituent quark can be reduced with the help of (\ref{pro30})
to a 4 point function 
\beq
  tr_{CD}[T_{j_\Gamma}(p',p)\Gamma'] =
    \int\!d^4\!x\,d^4\!z\;e^{ip'x-ipz}tr_{CD}[\langle 0|
    {\cal T} \psi(x)\bar\psi(0)\Gamma\psi(0)\bar\psi(z)|
    0\rangle\Gamma'] =
\eeq
$$
  = \int\!\dbar^4\!q\;\Pi_{\Gamma\Gamma'}(q-p,q-p',p,p')
$$
The polarisation functions $\Pi_{\Gamma\Gamma'}$  are calculated
are defined and calculated in the instanton liquid model in
\cite{Hut2} and other works.
For $\Gamma=\gamma_\mu\gamma_5$ the connected part of
the 4 point function is suppressed by $O(n_R^{1/2})$.
In leading order in the instanton density only the disconnected part
contributes and we get
\beq\label{pro33}
  T_{j_{\mu 5}^{GI}}(p',p) = S(p')\gamma_\mu\gamma_5 S(p)
\eeq
Inserting (\ref{pro33}) in (\ref{pro30}) and comparison with
(\ref{pro4}) leads to
\beq\label{pro34}
  \langle p's'|J_{\mu 5}^{GI}(0)|ps\rangle =
    \bar u_{s'}(p')\gamma_\mu\gamma_5 u_s(p)
\eeq
$$
    G_1^{GI}(q^2) = 1 \quad,\quad G_2^{GI}(q^2)=0
$$
Note, that $\Pi_{\Gamma\Gamma'}$ was calculated in singular gauge, but
the connected part is suppressed in any gauge and the disconnected
part only depends on the propagators, which cancel out anyway.
The form factors $G^{GI}_{1/2}(q)$ are indeed gauge invariant.
The result coincides with a model of free massive quarks.
Further we see that the current is not conserved. Conservation
depands $q^2G_2=MG_1$, which is clearly not satisfied by (\ref{pro34}).
In the one instanton approximation one can work from the very beginning
with the effective 't Hooft vertex \cite{tHo} which explicitly
breaks the $U(1)$ symmetry and therefore contains the anomaly.

The result for the GI form factors (\ref{pro34}), although not
consistent with the experimental value, is {\it up to now} at least
theoretical consistent.

\section{The Anomaly Form Factor $A(q)$ *}\label{sec67}

We will now calculate the anomaly form factor $A$.
Using again the reduction formula with insertion of the anomaly
current $B(0)=a(0)$ we have to calculate the 3 point function $T_a(p,s)$.
In the instanton model the field operator $a(0)$ is replaced by
a classical field $a_A(0)$ where $A=\sum_I A_I$ is a multi instanton
configuration inserted in $a$. In a given background $A$ the correlator
can be written in the form
\beq
  \langle 0|{\cal T}\psi(x)a(0)\bar\psi(z)|0\rangle_A =
  a_A(0)\langle 0|{\cal T}\psi(x)\bar\psi(z)|0\rangle_A =
  a_A(0)S_A(x,z)
\eeq
where $S_A(x,z)$ ist the quark propagator in the multi instanton
background $A$. The r.h.s. has now to be averaged over the collective
coordinates $\gamma_I$ of all instantons. Without the factor $a_A(0)$
this is just the averaged quark propagator calculated in \cite{Hut2}.
$a_A(y)$ is $2N_f$ times the topological charge density at spacetime
point y.
In the vicinity of an instanton of charge $Q_I=\pm 1$ the charge density
has a positive/negative bump and is small elsewhere. Therefore
$a_A(y)$ is only nonzero when there is at least one instanton near y.
Let us fix exactly one instanton in the vicinity of $y=0$. The
orientation and charge of the remaining instantons can be averaged
independently, but when averaging the locations $z_I$ the domain
near $y$ has to be avoided. The next step is to assume 2 instantons
near y and so on. The relative error we make by neglecting these further
contributions and by forgetting about the restriction on $z_I$ are
both of $O(n_R)$. In leading order in the instanton density we
can therefore fix one instanton near $y=0$ and take only this contribution
to $a_A(0)$ into account. The remaining instantons can be averaged
as in the pure propagator case and the diagrams which have to be summed
and averaged are the same except for the fixing of one instanton $I$.
The propagator consists of a chain of instanton scatterings $A_J$
($J=1\ldots N$).
Repeated scattering at this vertex is allowed.
There are two cases: The first case is that all instantons
left to all occurrences of instanton $I$ are different to all instantons
right to all occurences of instanton $I$.
In leading order in $1/N_c$ all instantons in the middle section
from the first up to the last occurence of $A_I$ are different to the
exterior instantons. The instantons
on the left and on the right can be averaged independently leading
to averaged multi-instanton propagators. Averaging the middle section,
but fixing $I$ leads to the effective vertex $M_I$.
The free part of the correlator in momentum space is therefore
$$
  T_a^{free}(p,s) = \langle 2N_fQ_I(z_I)\quad\makebox(22,2)
{\begin{picture}(22,5.85)
  \thicklines
  \put(11,2.85){\circle{6}}
  \put(11,2.85){\makebox(0,0)[cc]{$M_I$}}
  \put(0,2.85){\line(1,0){8}}
  \put(4,2.85){\makebox(0,0)[cc]{$<$}}
  \put(14,2.85){\line(1,0){8}}
  \put(18,2.85){\makebox(0,0)[cc]{$<$}}
  \put(3,1){\makebox(0,0)[cc]{$p$}}
  \put(17,1){\makebox(0,0)[cc]{$s$}}
 \end{picture}
}
\rangle_I =
$$
\beq\label{pro36}
  = -2iN_f\hat Q(p-s)\sqrt{M_pM_s}S(p)\gamma_5 S(s)
\eeq
$$
  Q_I(z_I)={1\over 2N_f}a_{A_I}(0)=\pm{6\over\pi^2}
    \left(\rho\over z_I^2+\rho^2\right)^4
$$
$Q_I(z_I)$ is the charge density of one instanton of charge $Q_I=\pm 1$
and $\hat Q(q)={1\over 2}(q\rho)^2K_2(q\rho)$
its fourier transform\footnote{
I apologize for the overload of the symbol $K$:
$K_2(q\rho)$ is a modified Bessel function, $K_\mu(x)$ is the
anomaly current and $K_{1/2}(q)$ its form factors.}
for $Q_I=+1$. For $p^2=s^2=M^2$
the term $\sqrt{M_pM_s}$ is just the onshell mass $M$. Inserting
(\ref{pro36}) into (\ref{pro30}) and comparision with (\ref{pro4})
we get for the free part of the anomaly form factor:
\beq
  A^{free}(q)=-N_f\hat Q(q) \quad,\quad A^{free}(0)=-N_f
\eeq
The second case is, that there are common instantons to the left
and to the right of instanton $I$. The connected part of the
correlator and the form factor are
\vspace{5mm}
\beq\label{pro38}
  T_a^{conn}(p,s) = \bigg\langle 2N_fQ_I(z_I) \makebox(25,2)
{\begin{picture}(25,21)
  \thicklines
  \put(3,5){\framebox(19,8)[cc]{$C^s$}}
  \put(12,18){\circle{6}}
  \put(12,18){\makebox(0,0)[cc]{$M_I$}}
  \put(7.50,15.50){\oval(3,5)[lt]}
  \put(17,15.50){\oval(4,5)[rt]}
  \put(6,15){\makebox(0,0)[cc]{$\vee$}}
  \put(19,15){\makebox(0,0)[cc]{$\wedge$}}
  \put(3,3.50){\oval(6,3)[rb]}
  \put(22,3.50){\oval(6,3)[lb]}
  \put(3,2){\makebox(0,0)[cc]{$<$}}
  \put(22,2){\makebox(0,0)[cc]{$<$}}
  \put(1,4){\makebox(0,0)[cc]{$p$}}
  \put(24,4){\makebox(0,0)[cc]{$s$}}
  \put(6,13){\line(0,1){3}}
  \put(9,18){\line(-1,0){2}}
  \put(19,13){\line(0,1){3}}
  \put(15,18){\line(1,0){2}}
  \put(19,5){\line(0,-1){2}}
  \put(6,5){\line(0,-1){2}}
  \put(25,2){\line(-1,0){4}}
  \put(0,2){\line(1,0){4}}
 \end{picture}
}
\bigg\rangle_I =
\eeq
\vspace{5mm}
$$
  = -4(N_f-1)i\hat Q(p-s)C_5^s(p-s)F_5(p-s)\sqrt{M_pM_s}
     S(p)\gamma_5 S(s)
$$
\beq
  A^{conn}(q) = -2(N_f-1)\hat Q(q)C_5^s(q)F_5(q) \quad,\quad
  A^{conn}(0) = N_f-1
\eeq
For one flavor the connected part is zero as it should.
For two flavors the result can easily derived by using the formulas of
\cite{Hut2}.
The total anomaly form factor for zero momentum transfer
\beq
  A(0)=A^{free}(0)+A^{conn}(0) = -1
\eeq
is independent of the number of flavors!
This result is welcomed due to the following
argument: The form factors of the axial singlet currents $j_{\mu 5}$
should not  depend on any quark flavor which is not involved in
the particle state. One expects that they are independent of $N_f$.
Due to (\ref{pro3}) matrix elements of $a(x)$ must then be
independent of $N_f$ too.
But this is not obvious because $a(x)$ is explicitly proportional to
$N_f$ and the gluonic field is not flavor sensitive. The calculation
given above shows how the quark interaction cancels the free part, which
is proportional to $N_f$, so that the total form factor is independent
of $N_f$ at least at zero momentum transfer.

\section{The Gluonic Form Factors $K_{1/2}^{GI}(q)$}\label{sec68}

Now we come to the calculation of $K_{1/2}(0)$. The previous calculation
can be copied with minor changes. $a(0)$ has to be replaced by
$K_\mu(0)$. This in turn induces the replacement
\beq
  2Q(z_I)    \leadsto G_\mu(z_I):={1\over N_f}K_{A_I}^\mu(0) \quad,\quad
  2\hat Q(q) \leadsto \hat G_\mu(q)
\eeq
$G_\mu(z_I)$ is $K_\mu(0)$ where the gauge field is an instanton
centered at $z_I$ of charge $Q_I=+1$ and $\hat G_\mu(q)$ is its
fourier transform. In regular gauge we get
\bqa
  G_\mu^{reg}(z) &=& {1\over N_f}K_{A_I^{reg}}^\mu(0) =
  - {z_\mu(z^2+3\rho^2)\over\pi^2(z^2+\rho^2)^3} \\
  \hat G_\mu^{reg}(q) &=& -iq_\mu\rho^2K_2(q\rho)
  \stackrel{q\to 0}{\longrightarrow}-2iq_\mu/q^2 \nonumber
\eqa
With this replacement in (\ref{pro36}) and (\ref{pro38})
and comparison with (\ref{pro4}) $K_{1/2}(0)$
can be extracted:
\beq
  K_1^{reg}(q)=0 \quad,\quad
  \lim_{q^2\to 0}{q^2\over 2M}K_2^{reg}(q)=1
\eeq
In singular gauge we get
\beq\label{pro44}
  G_\mu^{sing}(z) = G_\mu^{reg}(z)+{z_\mu\over\pi^2 z^4}
  \quad,\quad
  \hat G_\mu^{sing}(q) = \hat G_\mu^{reg}(q) + 2iq_\mu/q^2
    \stackrel{q\to 0}{\longrightarrow} 0
\eeq
$$
  K_1^{sing}(q)=0 \quad,\quad
  \lim_{q^2\to 0}{q^2\over 2M}K_2^{sing}(q)=0
$$

An apparent observation is, that the anomaly form factor
$K_2(q)$ is gauge dependent and recieves a massless pole in
regular gauge. The reason for this is the gauge
dependence of the anomaly current $K_\mu$ itself. One can show
that the forward matrix elements $K_{1/2}(0)$ are GI
for small gauge transformations. A gauge transformation is called
small, when it can be smoothly deformed into the unit transformation.
On the other hand the gauge transformation, which transforms
an instanton from regular gauge to one in singular gauge is large,
because the regular solution can not be smoothly deformed into a
singular one due to the singularity.

The next striking observation is that the relation
\beq\label{pro45}
  K_1(q)-{q\over 2M}K_2(q) = A(q)
\eeq
is violated in singular gauge as can be seen from (\ref{pro44})
\beq\label{pro46}
  K_1^{sing}(q)-{q\over 2M}K_2^{sing} \neq A(q)
\eeq
Surface terms are the origin of this violation. For
the derivation of (\ref{pro45}) one has assumed the vanishing of
surface terms. If one replaces the plane wave solution for the state
by a wave packet, the state and therefore the matrix elements
decrease sufficiently fast at spacial infinity and there are no
surface terms. A experimental state is always a more or less localized
wave packet rather than an exact plane wave. Therefore in regular gauge
there are no surface terms and
\beq\label{pro47}
  K_1^{reg}(q)-{q\over 2M}K_2^{reg} = A(q)
\eeq
is valid for all $q$.
In order to work in singular gauge we have to choose a space-time
manifold $I\!\!R^4\backslash\{0\}$ to exclude the unphysical singularity.
This small hole should not affect the physics at large distances.
Therefore all coordinate space intagrals are integrals over the
domain $I\!\!R^4\backslash B_\eps(0)$. Partial integration
can now lead to surface terms at zero.
The surface term is non-zero in the case of $G_\mu^{sing}$ as can be seen
from (\ref{pro44}). This is the reason for the inequality (\ref{pro46}).
It is surprising that not the slowly decaying regular gauge field
causes a surface term at infinity but the strong singularity
at the instanton centers in
singular gauge leads to surface terms and to a violation of (\ref{pro45}).

The following conclusions should be drawn:
\begin{enumerate}
\item not to consider gauge dependent objects like $K_{1/2}(q)$ at all or
\item save the relation (\ref{pro45}) by using regular gauge
      although this violates the philosophy of \cite{Hut3} or
\item modify relation (\ref{pro45}) by including the surface
      terms and be careful when performing partial integrations.
\end{enumerate}
In the following discussion we take position 2.

\section{Discussion}\label{sec69}

Comparing the results for the form factors
$\Delta\Sigma^{GI}=G_1^{GI}(0)$, $G_2^{GI}(0)$, $A(0)$,
$K_1(0)=K_1^{reg}(0)$ and $K_2(0)=K_2^{reg}(0)$ summarized in
the last row of table \ref{tabpro3} we clearly see that they
are in contradiction. It is not possible to determine the
remaining form factors in a way that they are consistent
with (\ref{pro5}) and (\ref{pro13}). The most obvious contradiction
is $\Delta\Sigma^{GI}\neq A(0)$. An opposite sign of the
anomaly would at least be theoretical consistent and would lead
to the naive expectations. The only candidate for this violation
of the axial ward identities is the neglection of the non-zeromodes.
All other approximations respect the symmetries of QCD
as discussed in \cite{Hut2}.

Forte \cite{For} has derived the relation $\Delta\Sigma+A(0)=0$ in
the instanton model in the case of one quark flavor in
quenched approximation and density expansion.
$\Delta\Sigma$ was identified with $\Delta\Sigma^c$
and $K_2(0)$ was assumed to be zero (although not explicitly stated).
Therefore $A(0)=K_1(0)$
and from (\ref{pro12}) one can arrive at the welcomed result
$\Delta\Sigma^{GI}=0$.

In section \ref{sec68} I have shown that the anomaly contributes to
$K_2$ and not to $K_1$. This is the first discrepancy.
Further, in section \ref{sec66} I have shown that $\Delta\Sigma$
has to be identified
with $\Delta\Sigma^{GI}$. This is the second discrepancy.
It may turn out that the inclusion of non-zeromodes
removes the discrepancies in a way, that leads to a
phenomenological welcomed small $\Delta\Sigma^{GI}$. In the
one instanton approximation the inclusion is managable and
has been performed by \cite{Geshkenbein} for the meson correlators.
The consistent extension to the instanton liquid and to the quark
form factors was not yet managable.

These problems might be compared to calculations of the $\eta'$
mass. A brute force method of calculating the axial singlet
meson correlator and extracting $m_{\eta'}$ by a spectal fit
is not successful too. More elaborate arguments, given in
\cite{Hut4} allowed a successful determination of $m_{\eta'}$.
The instanton model was only used as a motivation for
a selfdual model of QCD. Maybe the same model
is able to solve the proton spin problem rather than a
brute force calculation.

It might also be possible that the spin problem can not be solved
on the level of individual constituent quarks formed out of current
quarks but is connected with a strong interaction in the axial
singlet channel between different constituent quarks.
This possibility in connection with
instantons is discussed in \cite{Dorokhov}.


\stepcounter{section}
\addcontentsline{toc}{section}{\Alph{section} References}
\parskip=0ex plus 1ex minus 1ex

\end{document}